\documentstyle[11pt,epsfig]{article}
\begin{document}
\baselineskip 100pt
\renewcommand{\baselinestretch}{1.5}
\renewcommand{\arraystretch}{0.666666666}
{\large
\parskip.2in
\newcommand{\be}{\begin{equation}}
\newcommand{\ee}{\end{equation}}
\newcommand{\sg}{\sigma}
\newcommand{\br}{\bar}
\newcommand{\fr}{\frac}
\newcommand{\lm}{\lambda}
\newcommand{\ra}{\rightarrow}
\newcommand{\al}{\alpha}
\newcommand{\bt}{\beta}
\newcommand{\pr}{\partial}
\newcommand{\hs}{\hspace{5mm}}
\newcommand{\th}{\theta}
\newcommand{\up}{\upsilon}
\newcommand{\dg}{\dagger}
\newcommand{\ve}{\varepsilon}
\newcommand{\acc}{\\[3mm]}
\newcommand{\ie}{{\it ie }}
\newcommand{\sh}{\sharp}
\noindent

\hfill DTP\,98/15

\vskip .5truein

\begin{center}
{\bf THE INVERSE SPECTRAL THEORY FOR THE WARD EQUATION AND FOR THE
2+1 CHIRAL MODEL\footnote{To Appear in Communications in Applied
Analysis}}
\end{center}
\begin{center}
A. S. F{\small OKAS}$^1$ and T. A. I{\small OANNIDOU}$^2$\\
$^1${\sl Department of Mathematics, Imperial College,
 London SW7 2BZ, UK\\
\mbox{and}\\
$^1$  Institute for Nonlinear Studies, Clarkson University,
Potsdam, NY 13699-5815, USA}\\
$^2${\sl Department of Mathematical Sciences, University of
Durham, Durham DH1 3LE, UK}
\end{center}

\bigskip
{\bf ABSTRACT} 
We solve the Cauchy problem of the Ward model in light-cone coordinates
using the inverse spectral (scattering) method.
In particular we show that the solution can be constructed by solving a
$2\times 2$ local matrix Riemann-Hilbert problem which is uniquely defined
in terms of the initial data.
These results are also directly applicable to the $2+1$ Chiral model.


\section{INTRODUCTION}

We study the Cauchy problem for the Ward model in light-cone
coordinates:
\begin{eqnarray}
&&Q_{xt}=Q_{yy}+[Q_y,Q_x], \hs x, y\in R, \hs t\geq0,
\label{light}\\
&&Q(x,y,0)=Q_0(x,y),
\label{in}
\end{eqnarray}
where [ , ] denotes the usual matrix commutator,
$Q(x,y,t)$ is a traceless $2\times 2$ anti-Hermitian matrix and $Q_0(x,y)$
is a $2\times2$ anti-Hermitian traceless matrix decaying
sufficiently fast as $x^2+y^2 \ra \infty$.

We shall solve this problem using the so-called inverse spectral
(scattering) method. 
This method is based on the fact that equation (\ref{light}) is the
compatibility condition of the following Lax pair,
\begin{eqnarray}
\mu_y-k\mu_x-Q_x\mu&=&0,\label{KL}\\
\mu_t-k^2\mu_x-(kQ_x+Q_y)\mu&=&0\label{Kt}, \hs k \in C,
\end{eqnarray}
where $\mu(x,y,t,k)$ is a  $2\times2$ matrix.
The transformation
\be
x,\, y,\, t\hs \longrightarrow\hs \fr{t-y}{2},\, x, \,   \fr{t+y}{2},
\ee
maps equation (\ref{light}) to the Ward model \cite{W} in laboratory
coordinates.
The Cauchy problem in this model is defined by
\begin{eqnarray}
&&Q_{tt}-Q_{xx}-Q_{yy}=[Q_x,Q_t]-[Q_x,Q_y], \hs x,y\in
R, \hs t \geq 0, 
\label{lab}\\
 &&Q(x,y,0)=Q_{1}(x,y), \hs Q_t(x,y,0)=Q_{2}(x,y).
\label{condl}
\end{eqnarray}
This problem can be solved by using the fact that equation (\ref{lab})
possess the following Lax pair
\begin{eqnarray}
(k-\fr{1}{k}) \mu_x- 2 \mu_y-(Q_x+\fr{Q_t-Q_y}{k})\,
\mu&=&0,
\label{lax}\\
2 \mu_t-(k+\fr{1}{k}) \mu_x-(-Q_x+\fr{Q_t-Q_y}{k})\,\mu&=&0,\hs \hs
k\in C.
\label{t}
\end{eqnarray}
The Cauchy problem (\ref{lab}), (\ref{condl}), was studied in \cite{V}
using the Lax pair (\ref{lax}), (\ref{t}).

Here we study the Cauchy problem (\ref{light}), (\ref{in}), using the Lax
pair (\ref{KL}), (\ref{Kt}). 
We also make some remarks about the Cauchy problem (\ref{lab}),
(\ref{condl}).

We note that the transformations
\be
Q_y\doteq -J^{-1}J_t,\hs  \hs Q_x\doteq -J^{-1}J_y,
\ee
and
\be
Q_x\doteq -(J^{-1}J_t+J^{-1}J_y), \hs\hs Q_t-Q_y\doteq -J^{-1}J_x,
\ee
map equations (\ref{light}) and (\ref{lab}) to equations
\be
\left(J^{-1}J_y\right)_y=\left(J^{-1}J_t\right)_x
\label{lgc},
\ee
and
\be
\left(J^{-1}J_x\right)_x=\left(J^{-1}J_{(t+y)}\right)_{(t-y)}, 
\label{lc}
\ee
respectively.
Thus, our results are directly applicable to the solutions of the Cauchy
problem for equations (\ref{lgc}) and (\ref{lc}).
These equations are the 2+1 integrable chiral equations in light-cone and
laboratory coordinates, respectively.

In order to simplify the rigorous aspects of our formalism we first assume
that $Q_0(x,y)$ is a Schwartz function which is small in the following
sense
\be
\int_{R^2} \left|\widehat{Q}_0(\xi,y)\right| d\xi\,dy \ll 1,
\label{cond}
\ee
where $\widehat{Q}_0$ is the fourier transformation of $Q_0$ in the $x$
variable.
This assumption excludes soliton solutions. 
We then indicate how the formalism can be extended in the case that the
above assumption is violated.
In the case that $Q_0$ is sufficient small, the inverse spectral method
yields a solution of the Ward model in light-cone coordinates through the
following construction.

\newtheorem{theor}{Theorem}
\begin{theor}
Let $Q_0(x,y)$, $x,y \in R$ be a $2\times2$ anti-Hermitian traceless 
matrix which is a Schwartz function and which is small in the sense of
equation (\ref{cond}).

(i) Given $Q_0(x,y)$, define $\mu^+(x,y,k)$, $k \in C^+=\{k\in C:
\mbox{Im} \,k\geq  0\}$ and $\mu^-(x,y,k)$, $k \in C^-=\{k\in
C:\mbox{Im}\,k\leq 0\}$ as the $2\times2$ matrix valued functions which
are the unique solutions of the linear integral equations
\be
\!\!\!\!\mu^+(x,y,k)\!=\! I\!+\!\fr{1}{4 \pi}\!\!\left(\int_0^\infty
\!\!dp\!\! \int_{-\infty}^y \!\!\!\!dy'\!\!-\!\!\int_{-\infty}^0
\!\!\!\!dp\!\!\int_y^{\infty }\!\!dy' \right)\!\!\int_R
\!\!dx' e^{ip\left(x-x'+k(y-y')\right)}
Q_{0x'}(x',y')\mu^+(x',y',k),
\label{ms}
\ee
and
\be
\!\!\!\!\mu^-(x,y,k)\!=\!
I\!+\!\fr{1}{4\pi}\!\!\left(\int^0_{-\infty}
\!\!\!\!dp \!\!\int_{-\infty}^y\!\! \!\!dy'\!\!-\!\!\int^{\infty}_0\!\! dp
\!\!\int_y^\infty \!\!dy' \right)\!\!\int_R \!\!dx'
e^{ip\left(x-x'+k(y-y')\right)}
Q_{0x'}(x',y')
\mu^-(x',y',k), 
\label{mp}
\ee
where $I$ denotes the $2\times2$ unit matrix.

(ii) Given $\mu^\pm$ define the $2\times2$ matrix $S(x+ky,k)$, $x,y,k \in
R$, by
\begin{eqnarray}
\!\!I-S\!\!\!&=&\!\!\!\left(I-\fr{1}{4\pi}\int_{-\infty}^0
dp\,e^{ip(x+ky)}\, 
\int_{R^2} dx'\,dy'\,
e^{-ip\left(x'+ky'\right)}\,Q_{0x'}(x',y')\,\,\mu^+(x',y',k)
\right)^{-1}\!\times\nonumber \\
\!\!\!\!\!&&\!\!\left(I-\fr{1}{4\pi}\int_0^\infty dp\,e^{ip(x+ky)}\,
\int_{R^2} dx'\,dy'\,
e^{-ip\left(x'+ky'\right)}\,Q_{0x'}(x',y')\,\,\mu^-(x',y',k)
\right).
\label{scatt}
\end{eqnarray}

(iv) Given $S(x+ky,k)$ define the sectionally holomorphic function
$M(x,y,t,k)=M^+(x,y,t,k)$ for $k \in C^+$, $M(x,y,t,k)=M^-(x,y,t,k)$ for
$k \in C^-$ as the unique solution of the following $2\times2$
Riemann-Hilbert problem
\begin{eqnarray}
&&M^-(x,y,t,k)=M^+(x,y,t,k)\left(I-S(x+ky+k^2t,k)\right), \hs k\in
R,\label{RH}\\
&&\det M=1,\\
&&M=I+O(1/k), \hs k \ra \infty.
\end{eqnarray}

(v) Given $M(x,y,t,k)$ define $Q$ as
\be
Q_x (x,y,t)=\fr{1}{2i \pi} \int_R dk \,M^+(x,y,t,k) S(x,y,t,k).
\label{Q}
\ee
Then $Q$ solves equation (\ref{light}) and $Q(x,y,0)=Q_0(x,y)$.
\end{theor}

The rest of the paper is organized as follows. 
In section 2 we derive the main theorem. 
In section 3 we show how the relevant formalism can be extended to include
soliton solutions. 
In section 4 we briefly discuss the Cauchy problem for the Ward model in
laboratory coordinates.

We now make some remarks about related work.
A method for solving the Cauchy problem for decaying initial data for
integrable evolution equations in one spatial variable was discovered in
1967 \cite{GGKM}.
This method which we refer to as the inverse spectral method, reduces the
solution of the Cauchy problem to the solution of an inverse scattering
problem for an associated linear eigenvalue equation (namely for the
$x$-part of the associated Lax pair).
Such an integrable evolution equation in one spatial dimension is the
chiral equation; the associated $x$-part of the Lax pair is 
\be
\mu_x=-\fr{Q_x}{k}\mu,
\ee
where the eigenfunction $\mu(x,t,k)$ is a $2\times2$ matrix, $k$ is the
spectral parameter and $Q(x,t)$ is a solution of the chiral equation.

Each integrable evolution equation in one spatial dimension has several
two spatial dimensional integrable generalizations.
An integrable generalization of the chiral equation is (\ref{light}).
A method for solving the Cauchy problem for decaying initial data for
integrable evolution equations in two spatial variables appeared in the
early 1980 (see reviews \cite{BC},\cite{FS}).
For some equations such as the Kadomtsev-Petviashvili I equation, this
method is based on a
\underline{nonlocal Riemann-Hilbert} problem, while for other equations
such as
the Kadomtsev-Petviashvili II equation, this method is based on a  certain
generalization of the Riemann-Hilbert problem called the
\underline{$\bar{\partial}$ (DBAR)} problem.

It is interesting that although equation (\ref{light}) is an equation in
two spatial variables, the Cauchy problem can be solved by a local
Riemann-Hilbert problem.
This is a consequence of the fact that the equation (\ref{KL}) is a first
order ODE in the variable $x-ky$.

For integrable equations, there exist several different methods for
constructing exact solutions. Such exact solutions for the Ward model in
laboratory coordinates have been constructed in \cite{W,V}.
In particular, Ward constructed soliton solutions using the so-called
dressing method \cite{ZM}.
These solutions are obtained by assuming that $M(x,y,t,k)$ has simple
poles. 
In this case the corresponding solitons interact trivially, that is
they pass through each other without any phase-shift.
Recently, new soliton \cite{W1,I} and soliton-antisoliton solutions
\cite{I} were derived, by assuming that $M(x,y,t,k)$ has double or
higher order poles.
The corresponding lumps interact nontrivially, namely they
exhibit $\pi/N$ scattering between $N$ initial solitons.

The formalism presented in section 3 can also be used to obtain exact
soliton solutions. 
In particular, it is shown in section 3 that if the assumption
(\ref{cond}) is
violated then $M(x,y,t,k)$ still satisfies the Riemann-Hilbert problem
(\ref{RH}) but now it is generally a meromorphic as opposed to a
holomorphic function of $k$.
The solitonic part of the solution $Q(x,y,t)$ is generated by the
poles of $M$.
The main advantage of this approach is that it can be used to establish
the \underline{generic role}  played by the soliton solutions.
Namely, it is well known \cite{7} that the long time behaviour of the
solution of a local Riemann-Hilbert problem of the type (\ref{RH}) where
$M$ is a meromorphic
function of $k$, is dominated by the associated poles.
Thus the long time behaviour of $Q(x,y,t)$ with arbitrary
decaying initial data $Q_0(x,y)$ is given by the multisoliton solution.

\section{THE CAUCHY PROBLEM WITHOUT SOLITONS}

In this section we prove Theorem 1.

We first consider the direct problem, ie, we show that the spectral data
$S(x+ky,k)$ are well defined in terms of the initial data $Q_0(x,y)$.
Replacing $Q(x,y,t)$ by $Q_0(x,y)$ in equation (\ref{KL}) we find
\be
\fr{\pr \mu(x,y,k)}{\pr y}-k\fr{\pr \mu(x,y,k)}{\pr x}-Q_{0x}(x,y)\,
\mu(x,y,k)=0.
\label{f}
\ee
Let $\widehat{\mu}(p,y,k)$ denote the $x$-Fourier transform of
$\mu(x,y,k)$.
Then equation (\ref{f}) gives
\be
\widehat{\mu}_y-ipk\widehat{\mu}-\int_R
dx\,e^{-ipx}\,Q_{0x}(x,y)\,\,\mu(x,y,k)=0.    
\label{four}
\ee
Equations (\ref{ms}) and (\ref{mp}) are integrable forms of equation
(\ref{four}) with different initial values.
Under the small norm assumption (\ref{cond}), equations (\ref{ms}) and
(\ref{mp}) are uniquely solvable in the space of bounded continuous
functions $f(x,y)$ such that $f-I$ has a finite $L_1$ norm.
Since the dependence on $k$ of the kernel of the integral equations
(\ref{ms}) and (\ref{mp}) is analytic, the functions $\mu^\pm(x,y,k)$
are analytic in $k$ for $\pm Im\,k\geq 0$.

Equations (\ref{ms}) and (\ref{mp}) can also be written in the form
\be
\mu^\pm(x,y,k)=I+\int_{R^2} dx' dy' G^\pm(x-x',y-y',k)
\,Q_{0x'}(x',y')\,\,\mu^\pm(x',y',k),
\label{aut}
\ee
where
\be
G^\pm(x,y,k)=\fr{i}{4\pi^2}\int_{R^2} dp \,dl \,
\fr{e^{i(px+ly)}}{kp-l}I,\hs k\in
C^\pm.
\label{Green}
\ee
We note that $G^\pm$ can be evaluated in closed form,
\be
G^\pm(x,y,k)=\pm\fr{i}{2\pi  k
y}+\fr{\delta(y)}{2k}\left(\th(x)-\th(-x)\right),
\label{GT}
\ee
where $\delta(y)$ and $\th(x)$ denote the Dirac and the Heaviside
functions, respectively.
Indeed, writing $1/k=(k_R-ik_I)/|k^2|$ and using 
\be
\int_R dx \fr{e^{ipx}}{x+a+ib}=2\pi i \, (\mbox{sgn}x)
\,\theta(-xb)\,\,e^{ip(a+ib)}, \hs a, b, p \in R,
\ee
we find
\be
G^\pm(x,y,k)=-\fr{\mbox{sgn}x}{2\pi k}\int_R \! dl \,e^{ily}
\, \theta(xlk_I).
\ee
Recall that $G^+$ corresponds to $k_I\geq 0$; then in this case 
$(\mbox{sgn}x)\,
\theta(xlk_I)=\th (x) \th (l)- \th(-x) \th(-l)$, and the above equation
becomes
\begin{eqnarray}
\!\!G^+&=&-\fr{1}{2\pi
k}\left(\th(x)\int_0^\infty\!\! dl\,e^{il(y+i 0)}-\th(-x)\int_{-\infty}^0
\!\!dl\,e^{il(y-i 0)}\right)\nonumber\\
&=&\fr{i}{2\pi
k}\left(\fr{\th(x)}{y+i 0}+\fr{\th(-x)}{y-i 0}\right).
\end{eqnarray}
Using 
\be
\fr{1}{y\pm i 0}=\fr{1}{y}\mp\pi i \delta(y),
\ee
we find the expression for $G^+$ given by (\ref{GT}).
Similarly for $G^-$.

Using equation (\ref{GT}) it is straightforward to compute the large $k$
behaviour of $\mu^\pm$:
\be
\mu^\pm(x,y,k)=I\pm \fr{i}{2\pi k}\int_{R^2}
\!\!dx'\,dy'\,\fr{Q_{0x'}(x',y')}{y-y'}+\fr{1}{2k}\left(\int_{-\infty}^x
\!\!-\int^{\infty}_x\right)dx'Q_{0x'}(x',y)+O(\fr{1}{k^2}),
\label{as}
\ee
for $k \ra\infty$.
Thus
\be
\mu=I-\fr{Q_0(x,y)}{k}+O(\fr{1}{k^2}), \hs \hs
k \ra \infty.
\label{asy}
\ee

Taking the complex conjugate of equation (\ref{ms}), letting $p \ra -p$
and using the fact that
\be
\overline{Q}_{0_{11}}=-Q_{0_{11}}=Q_{0_{22}}, \hs \hs
\overline{Q}_{0_{12}}=-Q_{0_{21}},
\ee
we find
\begin{eqnarray}
&&\overline{\mu^+_{11}(x,y,\bar{k})}=\mu^-_{22}(x,y,k), \hs \hs\!\! \hs 
\overline{\mu^+_{21}(x,y,\bar{k})}= -\mu^-_{12}(x,y,k),\nonumber\\
&&\overline{\mu^+_{12}(x,y,\bar{k})}=-\mu^-_{21}(x,y,k),\hs \hs
\overline{\mu^+_{22}(x,y,\bar{k})}=\mu^-_{11}(x,y,k).
\label{sym}
\end{eqnarray}

Letting $\xi=x+ky$, $\eta=x-ky$, equation (\ref{f}) becomes
\be
2k \fr{\pr \mu}{\pr \eta}-Q_{0x}\, \mu=0.
\ee
Thus any two solutions of this equation are related by a matrix which is a
function of $x+ky$ and of $k$.
Hence
\be
\mu^-(x,y,k)=\mu^+(x,y,k)\left(I-S(x+ky,k)\right), \hs k \in R.
\label{rhk}
\ee
Equation (\ref{rhk}) and the symmetry relations (\ref{sym}), 
imply that
$I-S$ is a Hermitian matrix. 
In particular, the determinant of $I-S$ is real.
The determinant of equation (\ref{rhk}) yields
\be
\det \mu^-=\det \mu^+ \det(I-S).
\label{det}
\ee
Taking the complex conjugate of this equation and using the symmetry
relations (\ref{sym}), we find
\be 
\det \mu^+=\det \mu^- \det(I-S).
\label{det'}
\ee
Equations (\ref{det}) and (\ref{det'}) imply $\det(I-S)=\pm 1$.
However, equation (\ref{asy}) implies that
\be
\det \mu^\pm=1+O(1/k), \hs \hs k\ra \infty.
\label{detas}
\ee
Thus equation (\ref{det}) implies $\det (I-S)=1+O(1/k)$ as $k \ra \infty$,
and since $\det(I-S)=\pm1$ it follows that
\be
\det (I-S)=1.
\label{dets}
\ee
Equations (\ref{det}) and (\ref{dets}) imply
\be
\det \mu^+=\det \mu^-.
\label{detor}
\ee
Since $\mu^\pm$ are analytic in $C^\pm$, equations (\ref{detas}) and
(\ref{detor}) define a local Riemann-Hilbert problem \cite{AF}.
Its unique solution is 
\be
\det \mu^+=\det \mu^-=1.
\ee

Evaluating equation (\ref{rhk}) as $y \ra -\infty$ (keeping $x+ky$ fixed)
we find 
\be
I-S=\left(\lim_{y\ra -\infty} (\mu^+)^{-1}\right) \left(\lim_{y\ra
-\infty} 
(\mu^-)\right),
\ee
which is equation (\ref{scatt}).

We now consider the inverse problem, ie, we show how to construct the
solution of the
Cauchy problem (\ref{light}), (\ref{in}), starting from $S(x+ky,k)$.
Given $S(x+ky+k^2t,k)$, we define $M(x,y,t,k)$ as the solution of the
Riemann-Hilbert problem (\ref{RH}).
In general, if the $L_2$ norm with respect to $k$ of $S$ and of $\fr{\pr
S}{\pr k}$ are sufficiently small, then the problem has a unique solution.
However, in our particular case the solution exists without a small norm
assumption.
This is a consequence of the fact that $I-S$ is a Hermitian matrix.
Using this fact it can be shown (see for example \cite{FG})
that the homogeneous problem, ie, the problem
\begin{eqnarray}
&&\Phi^-=\Phi^+ (I-S), \hs \hs k \in R,\\
&&\Phi=O(\fr{1}{k}), \hs \hs  \hs \hs \, \,k \ra \infty,
\end{eqnarray}
has only the zero solution.

Given $M$, we define $Q(x,y,t)$ by equation (\ref{Q}).
A direct computation shows that if $M^+$ solves the Riemann-Hilbert
problem (\ref{RH}), ie, if $M^+$ satisfies
\be
M^+(x,y,t,k)=I+\fr{1}{2\pi
i}\int_Rdk'\fr{M^+(x,y,t,k')S(x+k'y+k'^2t,k')}{k'-(k+i0)}, \hs k \in R,
\label{cau}
\ee
and if $Q(x,y,t)$ is defined by equation (\ref{Q}) then $M^+$ satisfies
equations (\ref{KL}) and (\ref{Kt}).
Hence $Q$ satisfies equation (\ref{light}). 
Furthermore the investigation of the Riemann-Hilbert problem (\ref{RH})
at
$t=0$, implies that $Q(x,y,0)=Q_0(x,y)$.
Also since $I-S$ is Hermitian, $M^+$ and $M^-$ have the  proper
symmetry properties (see equations
(\ref{sym})), which in turn imply that $Q(x,y,t)$ is a
traceless anti-Hermitian matrix.

\underline{Remark} It is important to note that equation (\ref{KL})
involves $Q_x$ and not $Q$.
Because of this fact, equation (\ref{as}) reduces to equation (\ref{asy}).
This is to be contrasted
with the Kadomtsev-Petviashvili equation, whose Lax pair involves Q.
In that case equation (\ref{as}) simplifies  to equation (\ref{asy}) only
if
$\int_R Q(x,y) \,dx=0$.
This is the reason why it is usually assumed that the initial data of
the Kadomtsev-Petviashvili equation satisfy the above condition. 
Without this assumption, the inverse spectral method is more complicated
\cite{FS1}.

\section{SOLITON SOLUTIONS}

In this section we show how the formalism of section $2$ can be modified
to include the soliton solutions.

Since the matrix $I-S$ is Hermitian of  determinant one, it can be
represented as
\be
I-S=
\left(\begin{array}{llcl} 1 && \, \overline{\al}\\
\al && 1+|\al|^2
\end{array}\right),
\ee
where $\al$ is an arbitrary function of $(x+ky+k^2t,k)$.

Then equation (\ref{RH}) becomes
\be
(M_1^- \hs M_2^-)= (M_1^+ \hs M_2^+) \left(\begin{array}{llcl} 1 && \,
\overline{\al}\\
\al && 1+|\al|^2
\end{array}\right),
\ee
where $M_1^+$ and $M_2^-$ are 2-dimensional column vectors which are
functions of $(x,y,t,k)$.
In particular
\begin{eqnarray}
M_1^-&=&M_1^++\al\, M_2^+.
\label{M1}
\end{eqnarray}

Equations (\ref{ms}) and (\ref{mp}) are Fredholm integral equations of
the second type; thus they may have homogeneous solutions.
These homogeneous solutions which correspond to discrete
eigenvalues are rather important since they give rise to  solitons.
We assume that there exists a finite number of discrete eigenvalues and
that they are all simple.
Then Fredholm theory implies that $M^+_1$ admits the representation
\be
M_1^+=m_1^++\sum_{l=1}^N \fr{\phi_l(x,y,t)}{k-k_l},
\label{H1}
\ee
where $m_1^+(x,y,t,k)$ is analytic for $k\in C^+$ and the vectors
$\phi_l(x,y,t)$, $1\leq l \leq N$ are homogenous solutions of the first
column vector of equation (\ref{ms}).
Following the arguments of \cite{BC1} it can be shown that
\be
\phi_l(x,y,t)=-c_l(x+k_ly+k_l^2t)\, M_2^+(x,y,t,k_l),
\ee
where $c_l$ is a scalar function of the argument indicated.
Hence equation (\ref{H1}) becomes
\be
M_1^+(x,y,t,k)=m_1^+(x,y,t,k)-\sum_{l=1}^N \fr{c_l(x+k_l y+k_l^2 t)
\,M_2^+(x,y,t,k_l)}{k-k_l}.
\label{hom}
\ee
Substituting this equation  into equation (\ref{M1}) solving the resulting
Riemann-Hilbert problem we find 
\be 
\!\!\!\!M_1^-(x,y,t,k)+\sum_{l=1}^N\!
\fr{c_l(x+k_ly+k_l^2t)
\,M_2^+(x,y,t,k_l)}{k-k_l}\!=\!\!\left(\!\!\begin{array}{llcl} 1\\
0
\end{array}\!\!\right)\!-
\fr{1}{2i \pi} \!\!\int_R\!\! \fr{\al(x+k'y+k'^2t)
M_2^+(x,y,t,k')}{k'-(k-i
0)} dk'
\label{hom1}.
\ee
Let
\be
M_2^+(x,y,t,k)=\left(\begin{array}{llcl} A(x,y,t,k)\\ B(x,y,t,k)
\end{array}\right)
\label{pin}.
\ee
In what follows, for simplicity of notion we  suppress the $x,y,t$
dependence.
Using the notation (\ref{pin}) together with the symmetry relation
(\ref{sym}), equation (\ref{hom1}) becomes
\be
\left(\begin{array}{llcl} \overline{ \,\,\, \,\, B(\bar{k})} \\[1mm]
-\overline{A(\bar{k})}\end{array}\right)=
\left(\begin{array}{llcl}
1\\[1mm]0\end{array}\right)-\sum_{l=1}^N
\fr{c_l}{k-k_l}
\left(\begin{array}{llcl} A(k_l)\\[1mm] B(k_l)\end{array}\right)-
\fr{1}{2 i \pi}\int_R
\fr{dk'\,\al(k')}{k'-(k-i0)}\left(\begin{array}{llcl} A(k')\\[1mm] B(k')
\end{array}\right).
\label{sol}
\ee
Equation (\ref{sol}) express the solution of the Riemann-Hilbert problem
(\ref{RH}) in the case that solitons are included.

\underline{PURE SOLITONS}

Soliton solutions correspond to $\al=0$. In this case 
evaluating equation (\ref{sol}) at $k=\overline{k}_j$ we find
\begin{eqnarray}
\overline{ B(k_j)}&=&1-\sum_{l=1}^N \fr{c_l
}{\bar{k}_j-k_l}\,A(k_l),\nonumber\\
\overline{A(k_j)}&=&\sum_{l=1}^N \fr{c_l}{\bar{k}_j-k_l}\,B(k_l).
\label{ena}
\end{eqnarray}
The complex conjugate of these equations yields
\begin{eqnarray}
B(k_j)&=&1-\sum_{l=1}^N 
\fr{\overline{c}_l}{k_j-\bar{k}_l}\,\overline{A(k_l)},\nonumber\\
A(k_j)&=&\sum_{l=1}^N\fr{\overline{c}_l}{k_j-\bar{k}_l}\,\overline{
B(k_l)}.
\label{duo}
\end{eqnarray}
Equations (
\ref{ena}) and (\ref{duo}) determine $A(k_l)$ and $B(k_l)$,
$l=1, \dots ,N$.

Equation (\ref{hom1}) yields
\be
M_1^-=\left(\begin{array}{llcl} 1\\0\end{array}\right)-
\sum_{l=1}^N \fr{c_l}{k-k_l}\left(\begin{array}{llcl} A(k_l)\\
B(k_l)
\end{array}\right).
\ee
Thus using equation (\ref{asy}) we find
\be
Q_{11}=-Q_{22}= \sum_{l=1}^N c_l\,A(k_l),\hs \hs
Q_{21}=-\overline{Q_{12}}=\sum_{l=1}^N c_l\, B(k_l).
\label{soliton}
\ee

In summary, the $N$-soliton solution is given by equations
(\ref{soliton}), where $c_l=c_l(x+k_l y+k_l^2 t)$ and $A(k_l)$, $B(k_l)$
are the solutions of the equations (\ref{ena}) and (\ref{duo}).
In the case of 1-soliton equations (\ref{ena}) and (\ref{duo}) yield
\be
A(k_1)=-\fr{1}{1-\fr{|c_1|^2}{(\bar{k}_1-k_1)^2}} \fr{\overline{c}_1}
{\bar{k}_1-k_1}, 
\hs \hs
B(k_1)=\fr{1}{1-\fr{|c_1|^2}{(\bar{k}_1-k_1)^2}}.
\ee
Thus
\be
Q_{11}=-Q_{22}=-\fr{1}{1-\fr{|c_1|^2}{(\bar{k}_1-k_1)^2}}\fr{|c_1|^2}
{\bar{k}_1-k_1},
\hs \hs
Q_{21}=-\overline{Q_{12}}=\fr{c_1}{1-\fr{|c_1|^2}{(\bar{k}_1-k_1)^2}}.
\ee
Figure 1 represent a snapshot of the solution of equation (\ref{light}) by
taking $c_1=x+k_1y+k_1^2 t$ for  $k_1=i$ at time $t=-3$.

\begin{figure}[h]
\unitlength1cm
\hfil\begin{picture}(14,8)
\put(9,2.25){$x$}
\put(4,2.2){$y$}
\epsfxsize=13cm
\epsffile{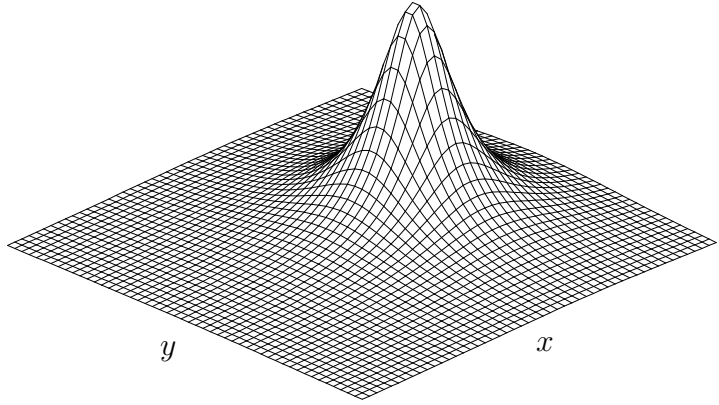}
\end{picture}
\caption{1-soliton solution of (\ref{light}) at $t=-3$.}
\end{figure}

\section{THE WARD MODEL IN LABORATORY COORDINATES}

Here we study the Cauchy problem (\ref{lab}), (\ref{condl}) using the Lax
pair (\ref{lax}), (\ref{t}).
In the case that $Q_1$ $Q_2$ are sufficient small, the inverse spectral
method yields a solution of the Ward model in laboratory coordinates
through the following construction.

\begin{theor}
Let $Q_1(x,y)$, $Q_2(x,y)$, $x,y \in R$ be $2\times2$ anti-Hermitian
traceless matrices which are Schwartz functions and which satisfy
the small norm conditions
\be
\int_{R^2} \left|\widehat{Q}_j(\xi,y)\right| d\xi\,dy \ll 1,
\hs \hs \hs j=1,2,
\ee
where $\widehat{Q}_j$ is the Fourier transformation of $Q_j$ in the $x$
variable. 

(i) Given $Q_1(x,y)$, $Q_2(x,y)$,  define $\mu^+(x,y,k)$, $k \in
C^+=\{k\in C:\mbox{Im} \,k\geq  0\}$ and $\mu^-(x,y,k)$, $k \in
C^-=\{k\in C:\mbox{Im}\,k\leq 0\}$ as the $2\times2$ matrix valued
functions which are the unique solutions of the linear integral equations
\be
\!\!\!\!\mu^+\!=\! I\!+\!\fr{1}{4 \pi}\!\!\left(\int_0^\infty
\!\!dp\!\! \int_{-\infty}^y \!\!\!\!dy'\!\!-\!\!\int_{-\infty}^0
\!\!\!\!dp\!\!\int_y^{\infty }\!\!dy' \right)\!\!\int_R
\!\!dx' e^{ip\left(x-x'+\fr{k^2-1}{2k}(y-y')\right)}
\left(Q_{1x'}+\fr{Q_{2}-Q_{1y'}}{k}\right)\mu^+,
\label{ms1}
\ee
and
\be
\!\!\!\!\mu^-\!=\!
I\!+\!\fr{1}{4\pi}\!\!\left(\int^0_{-\infty}
\!\!\!\!dp \!\!\int_{-\infty}^y\!\! \!\!dy'\!\!-\!\!\int^{\infty}_0\!\! dp
\!\!\int_y^\infty \!\!dy' \right)\!\!\int_R \!\!dx'
e^{ip\left(x-x'+\fr{k^2-1}{2k}(y-y')\right)}
\left(Q_{1x'}+\fr{Q_{2}-Q_{1y'}}{k}\right)\mu^-.
\label{mp1}
\ee
(ii) Given $\mu^\pm$ define the $2\times2$ matrix
$S(x+\fr{k^2-1}{2k}y,k)$, $x,y,k \in
R$, by
\begin{eqnarray}
\!\!I-S\!\!\!&=&\!\!\!\!\!\left(\!I-\fr{1}{4\pi}\int_{-\infty}^0
dp
\int_{R^2} dx'\,dy'\,
e^{ip\left(x-x'+\fr{k^2-1}{2k}(y-y')\right)}\,
\left(Q_{1x'}+\fr{Q_{2}-Q_{1y'}}{k}\right)\,\mu^+
\right)^{-1}\!\!\!\times \nonumber\\
\!\!\!&&\!\!\!\!\!\left(I-\fr{1}{4\pi}\int_0^\infty dp 
\int_{R^2} dx'\,dy'\,
e^{ip\left(x-x'+\fr{k^2-1}{2k}(y-y')\right)}\,
\left(Q_{1x'}+\fr{Q_{2}-Q_{1y'}}{k}\right)\,\mu^-
\right).
\end{eqnarray}

(iv) Given $S(x+\fr{k^2-1}{2k}y,k)$ define the sectionally
holomorphic
function
$M(x,y,t,k)=M^+(x,y,t,k)$ for $k \in C^+$, $M(x,y,t,k)=M^-(x,y,t,k)$ for
$k \in C^-$ as the unique solution of the following $2\times2$
Riemann-Hilbert problem
\begin{eqnarray}
&&M^-(x,y,t,k)=M^+(x,y,t,k)\left(I-S(x+\fr{k^2-1}{2k}y+\fr{k^2+1}{2k}t,k)
\right),
\hs k\in
R,\\
&&\det M=1,\\   
&&M=I+O(1/k), \hs k \ra \infty.
\end{eqnarray}

(v) Given $M(x,y,t,k)$ define $Q$ as
\be
Q_x (x,y,t)=\fr{1}{2i \pi} \int_R dk \,M^+(x,y,t,k) S(x,y,t,k).
\ee
Then $Q$ solves equation (\ref{lab}) and $Q(x,y,0)=Q_1(x,y)$,
$Q_t(x,y,0)=Q_2(x,y)$.
\end{theor}     

The proof of Theorem 2 is similar to section 2.
 
Equations (\ref{ms1}) and (\ref{mp1}) can also be written in the form  
\be
\mu^\pm(x,y,k)=I+\int_{R^2} dx' dy' G(x-x',y-y',k)
\left(Q_{1x'}+\fr{Q_{2}-Q_{1y'}}{k}\right)\,\,\mu^\pm(x',y',k),
\label{aut1}
\ee
where
\be
G^\pm(x,y,k)=\fr{1}{(2\pi)^2 i}\int_{R^2}dp\, dn \,
\fr{e^{i(px+ny)}}{(k-k^{-1})p-2n}I,
\hs k\in C^\pm,\\
\label{I}
\ee
or 
\be
G^\pm(x,y,k)=\pm\fr{1}{2\pi i k
y}+\fr{\delta(y)}{2k}\left(\th(x)-\th(-x)\right).
\ee
Substituting this equation into equation (\ref{aut1}), it is
straightforward to compute the large $k$ behaviour of $\mu^\pm$,
\be
\!\!\mu^\pm\!=\!I\pm\fr{i}{2\pi k}\int_{R^2}
\!\!dx'\,dy'\fr{\left(Q_{1x'}\!+\!\fr{Q_{2}-Q_{1y'}}{k}\right)}{y-y'}+
\fr{1}{2k}\left(\int_{-\infty}^x
\!\!-\int^{\infty}_x\right)dx'\left(Q_{1x'}\!+\!\fr{Q_{2}-Q_{1y'}}{k}\right)
+O(\fr{1}{k^2}),
\ee
for $k \ra\infty$.
Thus
\be
\mu=I-\fr{Q_1(x,y)}{k}+O(\fr{1}{k^2}), \hs \hs
k \ra \infty.
\ee

The corresponding soliton solutions of equation (\ref{lab}) can be derived
following the method of section 3.

\section{ACKNOWLEDGMENTS}

TI acknowledges support from EU ERBFMBICT950035.

\end{document}